\documentclass[11pt,onecolumn,amssymb,nofootinbib]{revtex4}
\usepackage{amsmath, amsthm, amscd, amssymb}
\usepackage{graphicx, braket}
\usepackage{bm}
\usepackage{bbm}
\usepackage{epstopdf}

\begin{document}

\title{\bf   The mechanism for creating   ``dynamical gravastar'' black hole mimickers  also explains formation of   ``little red dots''  } \bigskip

\author{Stephen L. Adler}
\email{adler@ias.edu} \affiliation{Institute for Advanced Study,
Einstein Drive, Princeton, NJ 08540, USA.}

\begin{abstract}
We argue that a high pressure phase transition of relativistic matter to a state with negative energy density, which leads to the formation of horizonless, globally unitary black hole mimickers,  also gives rise to the appearance of  ``little red dots''.  The energy source for the dots is the release of latent energy from the phase transition, and their excess redness is a result of this release taking place in a central region of exponentially small positive $g_{00}$, and hence very high gravitational redshift.
\end{abstract}

\vfill\eject

\maketitle

\leftline{Essay written for the Gravity  Research Foundation 2026 Awards for Essays on Gravitation}
\leftline{Submitted March 15, 2026}

\section{Introduction}

The aim of this essay is to relate two ideas that have appeared in the recent literature. The first idea, reviewed in \cite{adler1}, is the suggestion that horizonless black hole mimickers termed ``dynamical gravastars'' form when matter under extremely high pressure undergoes a phase transition to a state with  negative energy density, as is permitted when quantum effects are taken into account. These compact objects obey the ``strong'' and ``null'' energy conditions, and since they are globally unitary, they thus eliminate the ``information paradox'' problem associated with black hole horizons.

 The second idea, recently reviewed in \cite{Matthee}, is the suggestion  that the ``little red dots'' observed by the James Webb Space Telescope (JWST) are a new type of compact relativistic object,  arising from an accreting  ultramassive  black hole that is surrounded by an opaque ``atmosphere''.  In this proposal, the atmosphere  re-emits the energy released from accretion as very low energy, hence red,  radiation. This model has successes in explaining some observed spectral features, but since accreting black holes tend to have a highly time-variable luminosity, the model does not account for the observed lack of variability \cite{Zhang} in the luminosity of the little red dots. A new study by Ishikawa et al. \cite{ishi}  proposes a specific geometric picture, in which little red dots ``are composed of two key structures: a broad-line emitting, red continuum core (i.e.  the black hole central engine) and a narrow-line emitting, extended blue continuum (i.e. the host galaxy).''

 Our purpose  here is to  give an alternative mechanism to the ``atmosphere'' model for the formation of the ``dots'', with excess redness resulting from the large gravitational red shift accompanying generation of a dynamical gravastar. The energy release of the dots is attributed to the latent energy released in a first order phase transition, and  lack of high variability of the luminosity would be expected.

In Section 2 we briefly review the model for formation of black hole mimickers that we have developed in a series of papers \cite{adler1}.   The model is based on solving the Tolman-Oppenheimer-Volkoff  (TOV)  relativistic pressure balance equations, with an equation of state jump from a relativistic matter equation of state in the exterior, to an equation of state with positive pressure and negative matter energy density in the interior. As shown in Fig. 1,  at radii well above the equation of state jump, the metric component $g_{00}$ is already exponentially small, and thus there will be a large red shift when latent energy emitted below the jump radius exits to infinity.

In Section  3 we give a simplified model of a dynamical gravastar accreting matter from its environment, at a slow enough rate so that it grows by passing through the equilibrium states described in Section 2.  Following this growth using the Mathematica notebook for dynamical gravastars shows that as the mimicker mass grows, the depth of the gravitational potential well in which latent energy is released correspondingly grows, and so does the red shift of exiting radiation.   During mimicker  growth, it  continues to obey the mathematical black hole radius--mass relation (in geometrized units) $R \simeq 2M$, and has a photosphere with asymptotic  pressure $p(r) \propto {\rm constant} \times  e^{4M/r}$.

In Section 4 we pose some observational questions.  Finally, in Section  5 we summarize the multiple features that all follow from the single assumption of a high pressure phase transition of matter to a state with negative energy density.

\section{The dynamical gravastar model}

The dynamical gravastar model reviewed in \cite{adler1} consists of solving the TOV
equations \cite{oppen}--\cite{camen}  (which are equivalent to the Einstein equations, and are the relativistic generalization of the  hydrostatic equilibrium equations \cite{wein} used in nonrelativistic calculations  of stellar structure), assuming an equation of state undergoing a phase transition at very high fluid pressure. For a spherically symmetric fluid with pressure $p(r)$ and energy density $\rho(r)$, the TOV equations take the  general form
\begin{align}\label{newTOV}
\frac{dm(r)}{d r}=&4\pi r^2\rho(r)~~~,\cr
\frac{dp(r)}{d r}=&-\frac{\rho(r)+p(r)}{2} \frac{d\nu(r)}{d r} ~~~,\cr
\frac{d\nu(r)}{d r}=&\frac{N(r)}{1-2m(r)/r}~~~,\cr
N(r)=&(2/r^2)\big(m(r)+4\pi r^3 p(r)\big)~~~.\cr
\end{align}
Here  $m(r)$ is the volume integrated energy density within radius $r$,   and $\nu(r)=\log\big(g_{00}(r)\big)$.   The  general form TOV equations become a closed system when  supplemented by an equation of state $\rho(p)$ giving the energy density $\rho$ in terms of the pressure $p$.

In the model reviewed in
\cite{adler1} the exterior equation of state  is a relativistic matter equation of state $\rho(p)=3p$ for $p\leq {\rm pjump}$, and the interior equation of state is  $p + \rho(p) =\beta$, for $p > {\rm pjump}$, with $\beta$ a small positive constant that is a parameter of the model. When $\beta=0$ this is the Gliner \cite{gliner} equation of state, reviewed in \cite{adler1} and widely employed in the earlier black hole mimicker literature.   The constant $\beta$ is a stand-in for more complex equation of state physics, and its positive value maintains decrease of pressure from the center of the black hole mimicker to the exterior.  Although this interior equation of state is symmetric in the pressure $p$ and the energy  density $\rho$, the TOV equations are not symmetric, and they require that when all physical quantities are finite,  the pressure must  be continuous.  Hence the discontinuous jump in going from the exterior to the interior of the mimicker necessarily occurs in the energy density, and quantum renormalization effects have long been known \cite{renorm} to lead to negative energy densities.
The model exhibits a metric structure very similar to that of a Schwarzschild black hole in the exterior region, whereas  $g_{00}$ becomes exponentially small, but remains always positive, in the interior region.  This is illustrated in Fig. 1, which shows that the exponential decrease in $g_{00}$ commences  in the exterior region, at pressures well below  ${\rm pjump}$.

\section{A model for red dot formation  based on the structure of dynamical gravastar black hole mimickers}

In this section we propose a new mechanism for the formation of red dots.  Like the models reviewed in \cite{Matthee}, it is a variant on conventional black hole ideas.  But the redness in our proposal results not from absorption of energetic radiation by an opaque surround with subsequent re-emission  at red wavelengths, but rather from the gravitational redshift of initially energetic radiation as it climbs out of the deep gravitational potential well associated with the gravastar.  Moreover, in our proposal the source of the initial energetic radiation is not primarily gravitational energy released by the accretion of matter on a central black hole, but rather the latent energy released in the first order phase transition  to negative matter  energy density at pressures above pjump, as postulated in the dynamical gravastar model.

Consider an initially small dynamical gravastar embedded in a matter medium (likely hydrogen), from which it grows in mass by adiabatic accretion. When infalling matter equilibrates with the interior of the  dynamical gravastar,  our postulate implies that the ultra high pressure causes the matter to undergo a phase transition to a state of negative energy density, incrementally adding to the pre-existing gravastar interior negative energy density. This is accompanied by emission of latent energy as radiation that diffuses off to infinity, emerging highly red shifted, giving the observed ``red dot'' signature.   We assume the accretion is slow enough that the gravastar growth is quasi-equilibrium, in the sense that it can be approximated by a sequence of equilibrium structures described by the TOV equation Mathematica notebook of \cite{adler1}.   Because we are making a quasi-equilibrium calculation, we cannot estimate rate-dependent quantities such as the red dot luminosity.  This would require a computationally intensive time-dependent analysis requiring solution of partial differential equations, much harder than solution of the static TOV ordinary differential equations.

 The specific model for which we present results has $\beta=0.01$, and units of length chosen to make  pjump=1.  The sequence of structures is then generated by considering a sequence of central pressures $p(0)$ starting from just above pjump, and trending to the largest value for which we could get reliable computational results from the Mathematica notebook.  We will see that these central pressure values correspond to an increasing series of gravastar mass values $M$.

Sample results of this study are given in Table I. The columns of Table I, reading across, give the central pressure $p(0)$, the mass $M$ as seen at effective infinity, the radius ${\rm rjump}=r({\rm pjump})$ where the energy density jumps, the radius $r_{\rm MIN}$ at the kink minimum seen in the top panel of Fig.  1 which plots   $1-2m(r)/r$, the central value of $\log g_{00}(0)=\nu(0) $, and finally the jump value of  $\log g_{00}({\rm rjump})=\nu({\rm rjump}) $.  Scanning the table, one sees that (i) the mass $M$ increases monotonically, and very nonlinearly, with $p(0)$, (ii)  the gravastar nominal horizon at the kink minimum   obeys $r_{\rm MIN}\simeq 2 M$, and (iii) the central and jump radius values of $\log g_{00}(r)=\nu(r)$ decrease monotonically with increasing $M$, becoming very large negative for large $M$.  When $p(0)$ is not too close to ${\rm pjump}=1$, there is a sharp transition region from the exterior to the interior structure, with most of the transition taking place at radii above the jump radius, as seen in the stacked plots of Fig. 1.  This means that as the mass $M$ increases, the latent energy released inside the jump radius has an increasingly large redshift to exit to infinity, giving the observational presentation of an energetic ``red dot''.  Because the emitted radiation comes from a range of radii and reshifts, discrete spectral emission iines will be smeared out or broadened, which accords with the recent observations of \cite{ishi}.   Possibly this  could be used as a diagnostic to distinguish from models where redness comes from absorption by screening layers of matter, but this would require a detailed comparison with the alternative colliisional line broadening mechanism.

The pressure distribution in a dynamical gravastar has a tail extending to infinity, and the radius where the optical thickness of this tail is unity constitutes the ``photosphere'', giving the observed  radius.  An exact integration of the exterior region equations gives the formula
\begin{equation} \label{tail}
p(r)={\rm pjump}\,e^{2(\nu(\rm{rjump})-\nu(r))}=p(\infty) e^{-2 \nu(r)}~~~,
\end{equation}
and substituting the approximation (good except very near the jump radius) $g_{00}(r) \simeq 1-2M/r$ gives as a a good large radius fit to the tail  (as illustrated in Fig. 2)
\begin{equation} \label{tail1}
p(r)={\rm pjump}\,e^{2(\nu(\rm{rjump})+2M/r)}=p(\infty) e^{4M/r}~~~.
\end{equation}
A priori, one does not know the conversion from units of the Mathematica notebook to physical units, which would be needed to give a prediction of the observed dot size.

\section{Observational questions}

We note two questions that may be addressable in future observations.  First, is there an observed correlation between the dot radii and masses, and if so, can the luminosity tail be fitted with a simple functional form as in Eq. \eqref{tail1}?  Second, are there ways, other than analysis of line widths \cite{ishi} noted above,  of distinguishing observationally between redness arising from absorption of radiation by an opaque layer with subsequent thermalized re-radiation, and redness arising as a gravitational redshift effect?

\section{ Features following from the underlying assumption of a phase transition to a state with negative energy density}

An important criterion for judging the merit of a hypothesis in physics is to look for multiple  consequences stemming from a single assumption.  As reviewed in \cite{adler1} and as further elaborated in this essay,  our starting hypothesis of a high pressure first order phase transition of matter to a state with negative energy density has the following  consequences:

\begin{itemize}

\item{\bf No one-way horizon,  so no associated ``unitarity'' and ``information'' paradoxes}  In the dynamical gravastar model, the metric component $g_{00}$  is always positive down to the center of the compact object.  So physics remains globally unitary, and there is no associated ``black hole information paradox''. Globally the universe is causal, without the trillions of causally disconnected regions implied by the existence of horizons.
\item{\bf Mechanism for forming ``red dots''}  As elaborated in this essay, our starting hypothesis leads to a natural mechanism for the formation of the observed ``little red dots'' seen in the very early universe.
\item{\bf An active role for astrophysical black holes in galaxy formation.}   If   astrophysical black holes are   dynamical gravastars, since there is no horizon  they are ``leaky'' and can play an active role in galaxy formation, as proposed in \cite{adler2}, \cite{adler3} and \cite{silk}.
\item{\bf Interesting mathematics of autonomous differential equations}  As briefly discussed in \cite{adler1}, the scale invariant rewriting  of the exterior region TOV equations for our model takes the form of a two dimensional pair of autonomous differential equations, as previously noted in \cite{collins} and as discussed in more detail in \cite{kempfer} and \cite{adler4}.  This gives a powerful tool for analyzing the mathematical structure of the dynamical gravastar proposal.

\end{itemize}

\section{Acknowledgements}

I wish to thank Scott Tremaine for email correspondence emphasizing the role of the photosphere in determining the observed dot radius.  I also wish to thank Brent Doherty and Jessica Adler for proofreading the first draft, and the referee for asking about spectral implications from combining radiation from regions with different redshifts.

\section{Added Note}

After announcement of the 2026 Gravity Essay Competition results, I received an email from Dr. Julio Arrechea noting that in his thesis, and in papers with various collaborators culminating in the comparative article arXiv:2509.13421, he explored the role in black hole mimickers of a core component with negative energy density, arising from regularization in quantum theory of the stress-energy tensor.

\begin{table} [ht]

\caption{Survey of the model with $\beta=0.01$ and ${\rm pjump}=1$. Reading from left to right the columns give central pressure $p(0)$, mass   $M$ as seen at infinity, radius ${\rm rjump}=r({\rm pjump})$ (where we are taking ${\rm pjump}=1$), radius $r_{\rm MIN}$ at the kink minimum, central value of $\log g_{00}(0)=\nu(0) $, jump value of
$\log g_{00}({\rm rjump})=\nu({\rm rjump}) $.  The nonlinearity of the solution and the limiting value of $p(0$ are intrinsic features of the model, not computational artifacts.  For a start to a detailed study of nonlinear features of the model, see \cite{kempfer} and \cite{adler4}.}
\begin{tabular}{c c c  c c c}
\hline\hline
$p(0)$& $M$& ${\rm rjump}$& $r_{\rm MIN}$&  $\nu(0)$ & $\nu({\rm rjump})$ \\
\hline
1.01& 0.55  &0.862   & 1.1  & -7.7 & -5.7  \\
1.04& 10.85  &17.8   &21.7  &-19.7  &-11.7   \\
1.07 & 208.8  &344  & 418 & -31.6 & -17.6  \\
1.1 &4025  &6621   &8050  &-43.5  &-23.5   \\
1.13  &77,653   &127,750  &155,300  &-55.4  &-29.4   \\
1.206  &$1.356 \times 10^8$  &$2.316 \times 10^8$   &$2.815 \times 10^8$  & -85.5 & -44.3  \\
\hline\hline
\end{tabular}
\label{tab2}
\end{table}


\newpage
\vfill{\hfill}
\begin{figure}[b]
\begin{centering}
 \includegraphics[natwidth=\textwidth,natheight=300,scale=0.7]{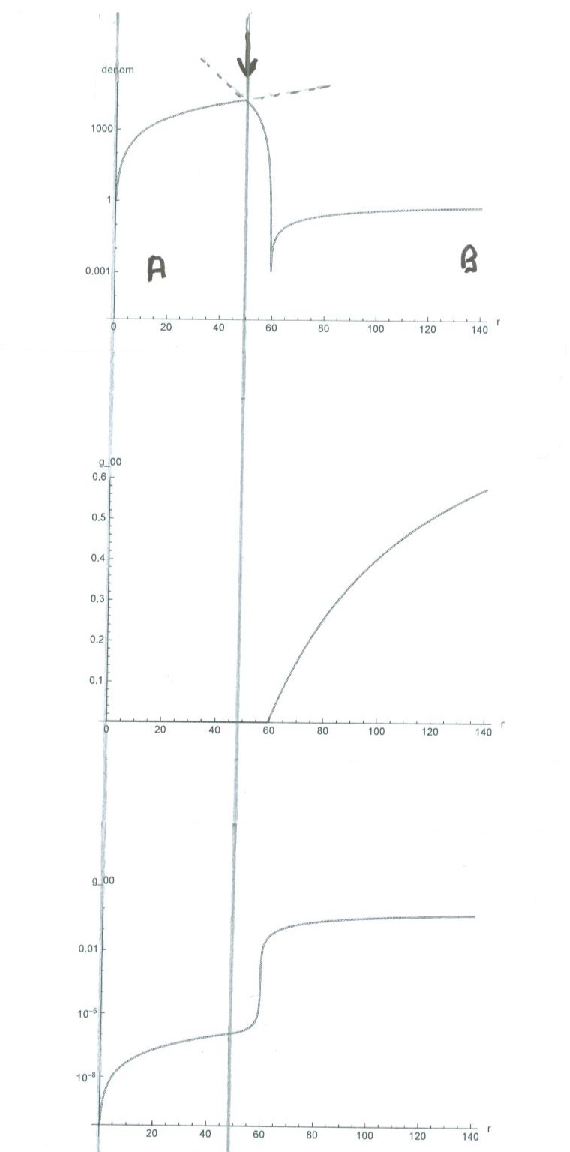}
\caption{This figure, taken from \cite{adler1}, was computed with $\beta=0.01$, central pressure $p(0)=1$, and ${\rm pjump}=0.95.$  In the figure plots of ${\rm denom}=1-2 m(r)/r$, together with  $g_{00}(r)$ on linear and logarithmic scales,  are stacked with their horizontal axes aligned.  The vertical line at $r=48.895$, marked by a down-pointing arrow, is where the discontinuity in equations of state at ${\rm pjump}=0.95$ appears; to the left of this line, in region A, the equation of state is $p+\rho=0.01$, and to the right of this line, in region B, the equation of state is $\rho=3p$.  The discontinuity in slopes of denom in the top panel shows up as the angle between the dashed lines being less than $\pi$, and arises because $m'(r)$ is not continuous where the equation of state is discontinuous.  The middle panel shows that at the cusp the metric component $g_{00}$ has nearly vanished on a linear plot, but the bottom panel shows that $g_{00}$ remains strictly positive down to zero radius, but precipitously drops to exponentially small values at radii below the radius $r\simeq 60$ of the cusp in denom. This gives a ``simulated horizon'', or perhaps better termed a ``mock horizon'', at the radius of the cusp.  Outside this radius the metric is very close to that of a Schwarzschild solution, while inside this radius the behavior is very different, with $g_{00}$ never going negative.}
\end{centering}
\end{figure}

\begin{figure}[b]
\begin{centering}
\includegraphics[natwidth=\textwidth,natheight=300,scale=1.0]{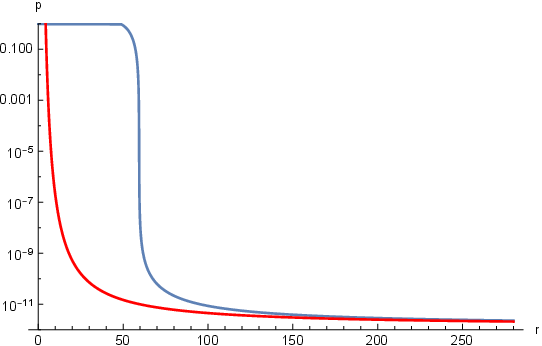}
\caption{An example  comparing  the tail formula of Eq. \eqref{tail1} (red curve) with the exact $p(r)$ computed from the TOV equations (blue curve).  The parameter values for this figure are those of Fig. 1, except that we have extended rmax  from 140 to 280 to give better coverage of the tail region.   }
\end{centering}
\end{figure}

\end{document}